\documentclass{article}

\usepackage{arxiv}

\usepackage[utf8]{inputenc} 
\usepackage[T1]{fontenc}    
\usepackage{hyperref}       
\usepackage{url}            
\usepackage{booktabs}       
\usepackage{amsfonts}       
\usepackage{nicefrac}       
\usepackage{microtype}      
\usepackage{lipsum}         
\usepackage{graphicx}
\usepackage{doi}
\usepackage{mathtools}
\usepackage[table,xcdraw]{xcolor}
\usepackage{multirow}
\usepackage{amsmath}
\usepackage{cleveref}       
\usepackage{bbm}
\usepackage{algorithm}
\usepackage{algorithmic}

\def\methodname{MDLT}
\def\fullmethodname{Music to Dance as Language Translation}

\title{Music to Dance as Language Translation using Sequence Models}

\date{}

\author{ \hspace{1mm}André Correia \\
	Universidade da Beira Interior and NOVA LINCS \\
	Covilhã, Portugal \\
	\texttt{andre.correia@ubi.pt} \\
	\And
	\hspace{1mm}Luís A. Alexandre \\
	Universidade da Beira Interior and NOVA LINCS \\
	Covilhã, Portugal \\
	\texttt{luis.alexandre@ubi.pt} \\
}

\renewcommand{\shorttitle}{DEFENDER}

\hypersetup{
pdftitle={DEFENDER: DTW-Based Episode Filtering Using Demonstrations for Enhancing RL Safety},
pdfsubject={q-bio.NC, q-bio.QM},
pdfauthor={André Correia, Luís A. Alexandre},
pdfkeywords={Demonstration, Reinforcement, Learning, Safety, Safe},
}

\begin{document}
\maketitle

\begin{abstract}
Synthesising appropriate choreographies from music remains an open problem. We introduce \methodname{}, a novel approach that frames the choreography generation problem as a translation task. Our method leverages an existing data set to learn to translate sequences of audio into corresponding dance poses. We present two variants of \methodname{}: one utilising the Transformer architecture and the other employing the Mamba architecture. We train our method on AIST++ and PhantomDance data sets to teach a robotic arm to dance, but our method can be applied to a full humanoid robot. Evaluation metrics, including Average Joint Error and Fréchet Inception Distance, consistently demonstrate that, when given a piece of music, \methodname{} excels at producing realistic and high-quality choreography. The code can be found at github.com/meowatthemoon/MDLT.
\end{abstract}


\section{Introduction}
\label{introduction}

Dance is an aesthetic performing art, that serves as a medium to convey intricate emotional nuances through choreography closely intertwined with music. The task of automatically generating choreographies from music has garnered significant interest within the research community, promising practical applications in various real-world domains like games and films \cite{aichoreographer,learn2dance,rhythm}. The current methodologies involving sequence posing, hand animation, or motion capture are not only laborious but also expensive.
However, the challenge of dance generation lies in balancing the inherent human creativity in a choreography with the need for the movements to remain coherent with the rhythm and genre of the music. 

\begin{figure}[!tb]
\centering
  \begin{minipage}{\linewidth}
    \makebox[\linewidth]{\includegraphics[width=0.7\linewidth]{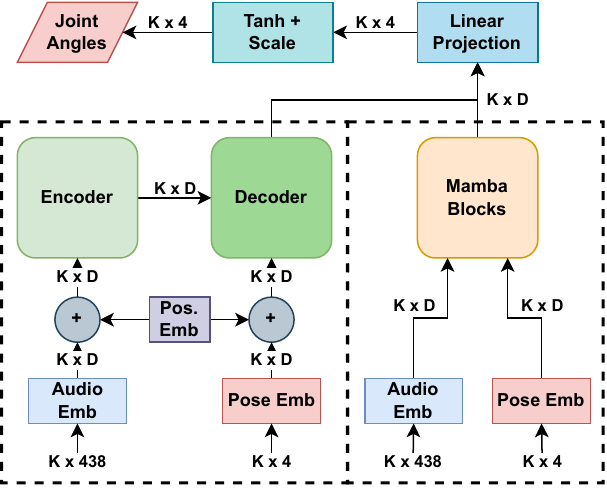}}%
  \end{minipage}
  \caption{Architecture of \methodname{} model variants. The audio features and poses first pass through their respective embedding layers. In the case of the Transformer variant (dashed left block) these embeddings are augmented with positional encoding. The encoder of the Transformer is conditioned on the audio features. The decoder of the Transformer is conditioned on the output of the encoder as well as the poses. The Mamba variant (dashed right block) receives both the audio and pose vectors. The embeddings of the final Mamba or Decoder block are projected to the pose dimensions. Finally, these values are activated using tanh and scaled to the joint angle range to produce the next poses. Only one of the dashed blocks is used.}
  \label{fig:diagram}
\end{figure}

Our work is grounded in two fundamental concepts that underpin our research. Firstly, we recognise the temporal coherence between the music piece and the dance choreography, emphasising that treating each component separately is not feasible. Secondly, we capitalise on the understanding that music and dance constitute unique languages with an inherent correspondence. In light of this, we propose to conceptualise the automatic dance generation challenge as a translation problem—akin to converting from the source language of music to the target language of dancing.
Embracing this perspective enables us to leverage the capabilities of sequence models \cite{attention,Mamba}, specifically designed for processing sequential data, and well-established in achieving success in language translation.

We introduce our method as \fullmethodname{} (\methodname{}). 
\methodname{} leverages a pre-existing data set of music-dance pairs and learns the mapping between both languages. The initial stages involve extracting musical features and joint angles from the data set. 
We propose to variants: \methodname-T using the Transformer architecture and \methodname-M using the Mamba architecture. The model is conditioned on the sequence of musical features and predicts the subsequent pose in the choreography. This predicted pose is incorporated into the choreography sequence, and the musical sequence advances in time. This iterative process continues until the culmination of the song.

We conducted a series of experiments using a subset of the AIST++ \cite{aichoreographer} data set and the entire PhantomDance \cite{danceformer} data set. Our method's performance was assessed across distinct music genres as well as collectively across all genres. Because the evaluation was performed using unseen music data, we show that our method learns to translate music into dance in a coherent and adaptive manner. The contributions of this paper can be summarised as follows:

\begin{itemize}
    \item We propose a new perspective to model the music conditioned dance generation as a translation task, where an agent learns the mapping between the music and dance languages from music-dance pairs in a data set. To our knowledge, this is the first study to propose this perspective and demonstrate its viability.
    
    \item We propose two variants: \methodname{}-T employs a modified Transformer architecture and \methodname{}-M employs the Mamba architecture.
    
    \item We compare the performance the two variants on a section of the AIST++ data set and the full PhantomDance data set.
    
    \item Experiments with a UR3 show that \methodname{}, can learn the mapping from audio to dance of different music pieces from different genres. \\Code and models are available at:github.com/meowatthemoon/MDLT.
\end{itemize}

\section{Related Work}
\label{sec:relatedwork}

\begin{table}[t]
\centering
\caption{Comparison of music features used by music-to-dance generation methods. Most methods use Librosa for music feature extraction. The most commonly used features are MFCC, MFCC delta, constant-Q chromagram, tempogram, and onset strength.}
\label{tab:features}
\begin{tabular}{c|ccccccc}
\hline
\textbf{Method}   & \rotatebox[origin=c]{90}{\textbf{Chromagram}} & \rotatebox[origin=c]{90}{\textbf{MFCC}} & \rotatebox[origin=c]{90}{\textbf{MFCC Delta}} & \rotatebox[origin=c]{90}{\textbf{Mel Spectogram}} & \rotatebox[origin=c]{90}{\textbf{Onset Strength}} & \rotatebox[origin=c]{90}{\textbf{Tempogram}} & \rotatebox[origin=c]{90}{\textbf{Others}} \\ \hline
AI Choreographer \cite{aichoreographer}  & x                   & x             &                     &                         &                         &                    & x               \\
Bailando \cite{bailando}         & x                   & x             & x                   &                         & x                       & x                  &                 \\
Choreomaster \cite{choreomaster}     &                     &               &                     & x                       &                         &                    &                 \\
DanceHat \cite{dancehat}          & x                   &               & x                   &                         & x                       & x                  & x               \\
DanceFormer \cite{danceformer}       & x                   & x             &                     &                         &                         &                    &                 \\
Dance with Melody \cite{dancewithmelody} & x                   & x             &                     &                         &                         &                    &                 \\
DeepDance \cite{deepdance}        &                     & x             & x                   & x                       &                         &                    & x               \\
M2C \cite{m2c}               & x                   & x             & x                   &                         & x                       & x                  &                 \\ \hline
\end{tabular}
\end{table}

Audio processing for dance generation encompasses various methodologies that leverage distinct techniques for feature extraction and sequence modelling. The most common practice is to use the Librosa \cite{librosa} library to extract audio features, where the different methods mostly disagree on the sampling rate and the types of features employed to represent audio vectors. Ablations in \cite{m2c} determined the more useful audio features for dance generation and emphasised the significance of their normalisation in the process. In contrast, the method in \cite{edge} diverges from using Librosa, utilising features extracted from the Jukebox \cite{jukebox} pre-trained model.

Early attempts at automatic audio-to-dance conversion using deep learning, utilised LSTM networks for sequence modelling \cite{dancewithmelody} and realised a sole LSTM could not converge. 
Instead of using LSTMs, in \cite{choreomaster}, the authors train two embedding spaces, where dances or music from the same style or genre, respectively, are mapped to similar embedding vectors.
A CNN then receives style and rhythm embedding to generate dance.

GANs \cite{gan} have also been employed for training generators conditioned on music to produce quality dances. In \cite{dancehat} the method incorporates stability concerns by augmenting the discriminator to penalise unstable poses. The method in \cite{deepdance} also utilises GANs but structures the architectures with LSTM layers to incorporate sequentially.

The advent of Transformer models \cite{attention} revolutionised sequence modelling, offering parallelised input processing and a masked attention mechanism for inter-sequence element contributions. 
The method in \cite{danceformer} combines adversarial learning with Transformer models. It employs two Transformer models, one conditioned on music to predict the correct pose on the beats, another conditioned on consecutive poses to predict the motion curve parameters between the two poses. This paper also introduced the PhantomDance data set.
In \cite{aichoreographer} the authors present AIST++ data set, an enhanced version of AIST with annotations of the dancer's poses. Additionally, they employed a three-Transformer framework that predicts the remaining half of the choreography given an audio sequence and initial choreography. 
The authors of \cite{bailando} deal with the dimensionality and precision dance pose generation issues by discretising the poses into two finite codebooks. These codebooks are estimated using variational auto-encoders. A Transformer model then learns the mapping between the audio sequences to a pose in each codebook.


While Transformers, with their self-attention mechanism, have demonstrated remarkable capabilities, their scalability is hindered by quadratic scaling concerning the size of the context window. Enter structured state space sequence models (SSMs) \cite{ssms}, which garnered attention for their linear scalability with sequence length. Particularly the Mamba \cite{Mamba} architecture, that aimed to combined the context-dependent reasoning of Transformers with the linear scalability of SSMs by introducing its selection mechanism. Mamba has surpassed Transformers in many sequence processing tasks. Because of this, we propose the variant \methodname{}-M using the Mamba architecture.

\section{Preliminaries}

\subsection{Transformers}

The main component of a Transformer \cite{attention} is the multi-head attention-layer which creates attention scores from input query \textbf{Q}, key \textbf{K} and value \textbf{V}. The attention module repeats its computations multiple times in parallel, by splitting the inputs passing them through N-heads.

\begin{equation}
\begin{aligned}
  Attention(\textbf{Q}, \textbf{K}, \textbf{V}, \textbf{M}) = softmax\left(\frac{\textbf{Q} \textbf{K} ^ T + \textbf{M}}{\sqrt{D}}\right)\textbf{V}
\end{aligned}
\end{equation}

where D is the number of channels in the attention layer and \textbf{M} is the attention mask. All head calculations are then combined together to produce a final Attention score. The mask parameter determines which other elements in the sequence can each element see to determine the attention score. In translation tasks, the mask of the encoder is a full attention mask where only the padded elements are hidden from the attention computation. The mask of the decoder is a causal mask, a upper triangular look-ahead mask where each element can only look at past elements.

\subsection{Structured State Space Sequence Models}

SSMs and Mamba are built upon the concept of continuous systems that maps a 1-D function $x(t) \xrightarrow{} y(t) \in \mathbb{R}$ through a hidden state $h(t) \in \mathbb{R}^N$. This process can be represented as a linear Ordinary Differential Equation (ODE):

\begin{equation}
\begin{aligned}
  h'(t) = \textbf{A}h(t) + \textbf{B}x(t),\ \ \ \ y(t) = \textbf{C}h(t)
\end{aligned}
\end{equation}

where, $\textbf{A} \in \mathbb{R}^{N \times N}$ serves as the evolution parameter, while $\textbf{B} \in \mathbb{R}^{N \times 1}$ and $\textbf{C} \in \mathbb{R}^{N \times 1}$ act as the projection parameters.

S4 adapt continuous systems for deep learning applications through discretisation by introducing a timescale parameter, $\Delta$, and converting the continuous parameters $\textbf{A}$ and $\textbf{B}$ into discrete parameters $\bar{\textbf{A}}$ and $\bar{\textbf{B}}$. This transformation is achieved using the zero-order hold (ZOH) method:

\begin{equation}
\begin{aligned}
  \bar{\textbf{A}} = \exp(\Delta\textbf{A}),  \ \ \ \
  \bar{\textbf{B}} = (\Delta\textbf{A})^{-1}(\exp(\Delta\textbf{A}) - \textbf{I}).\Delta\textbf{B}
\end{aligned}
\end{equation}

After the discretisation, the models can be rewritten as:

\begin{equation}
\begin{aligned}
  h'(t) = \bar{\textbf{A}}h(t) + \bar{\textbf{B}}x(t),  \ \ \ \ y(t) = \textbf{C}h(t)
\end{aligned}
\end{equation}

Lastly, the models compute the output through a global convolution:

\begin{equation}
\begin{aligned}
  \bar{\textbf{K}} = (\textbf{C}\bar{\textbf{B}}, \textbf{C}\bar{\textbf{A}}\bar{\textbf{B}}, ..., \textbf{C}\bar{\textbf{A}}^{K-1}\bar{\textbf{B}}),\ \ \ \ y(t) = x * \bar{\textbf{K}}
\end{aligned}
\end{equation}

where $\bar{\textbf{K}} \in \mathbb{R}^K$ represents a structured convolutional kernel, and $K$ denotes the length of the input sequence $x$.

\subsection{Translation}

In language modelling, the objective is to estimate the likelihood of a word given a preceding sequence of words. For a given source sentence $X = (x_1, x_2, ..., X_{N_s})$ and a corresponding target sentence $Y = (y_1, y_2, ..., y_{N_t})$, the translation model aims to sequentially predict the words in the target sentence. Here, $N_s$ and $N_t$ denote the lengths of the source and target sentences, respectively. More formally, the model should provide estimates of the conditional distribution $p(y_i \mid X, y_{1:i-1})$. 

In the context of our problem, we conceptualise the sequences of audio features as the source language and the sequences of dance poses as the target language. The goal is to estimate the parameters of a sequence model that best capture the mapping between these two languages. 

\section{Data Preparation}
\label{sec:pipeline}

\begin{figure*}[t]
\centering
  \begin{minipage}{\linewidth}
    \makebox[\linewidth]{\includegraphics[width=0.98\linewidth]{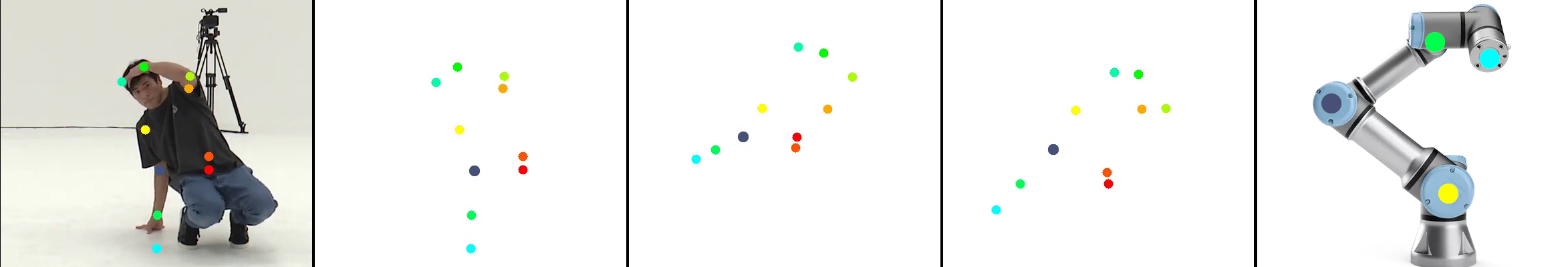}}%
  \end{minipage}
  \caption{Joint angle extraction from keypoints: First we obtain the arm and pelvis keypoints. Then we align the shoulders with the ground plane. Next, we align the spine vertically. Lastly, we extract the joint angles from the angles between the arm vectors.}
  \label{fig:keypoints}
\end{figure*}

\methodname{} learns the mapping between music pieces and dance choregraphies. To enable the model to interpret music, audio features are initially extracted from the music pieces, forming feature vectors that serve as the model's input. Then, because in this work we apply the method to teach a UR3 robot to dance, the human poses in the choreographies must be translated into the corresponding joint angles of the robot. Subsequently, the sequences of audio features and joint angles need to be synchronised into pairs, allowing the model to learn the mapping between music and the corresponding joint angles. The subsequent sections explain each data processing step in more detail.

\subsection{Data sets}

We use the AIST++ \cite{aichoreographer} and PhantomDance \cite{danceformer} data sets. Both provide music-to-dance pairs. AIST++ was built by reconstructing 3D poses from 2D multi-view videos sourced from the original AIST Dance Database. Comprising 1408 sequences executed by multiple dancers across 10 diverse dance genres. 
While multiple choreographies are available for each music piece in the data set, our focus on estimating the mapping from music to dance requires a one-to-one correspondence between audio and poses. Consequently, we assign a single choreography to each music piece. We amalgamate the music pieces from the original train, test, and validation splits, totalling 60 pieces with 6 pieces per genre. 
In the experiments section, we assign custom splits of the data set.

The PhantomDance data set was built by professional animators and is composed of 260 dance videos from over 100 different subjects.
Formally, the data sets can be expressed as:

\begin{equation}
\begin{aligned}
  \mathcal{D}' = \bigcup_{i=1}^{G} \{(M_{ij}, D_{ij})\}_{j=1}^{G_i}
\end{aligned}
\end{equation}

where $G$ is the set of music-genres, and $G_i$ is the number of pairs in the genre.

\subsection{Audio Features}

We extracted musical features from the audio files, thereby transforming the music into interpretable vectors for our model. Following a prevalent approach employed in the majority of music-to-dance studies, excluding \cite{edge}, we utilise the publicly available audio processing toolbox, Librosa, for feature extraction. The diversity in feature selection across various works prompted us, in synergy with the contributions of \cite{m2c}, to delve into different feature combinations explored by existing music-to-dance methodologies. A comprehensive overview of these features and their respective utilisation is provided in Table \ref{tab:features}. Drawing from this study, we used the MFCC, MFCC delta, constant-Q chromagram, tempogram, and onset strength features. These features are concatenated to form a 438-dimensional vector. Feature extraction is conducted at a rate of 60 times per second. Acknowledging the cautionary note of \cite{m2c},
 we normalise the features across each index of the 438 dimensions, constraining the values within the range of 0.1 to 0.9. 

\subsection{Joint Angles}

\methodname{} learns to translate music feature sequences into corresponding pose sequences. In our application, involving a UR3 robotic arm, the human poses from our data set need to be converted to corresponding joint angles.
While our current application involves translating human poses into the corresponding joint angles for the UR3 robot, it's important to highlight that our method is not limited to this particular robot or joint structure.
Our method can be extended to other robotic platforms by adjusting the mapping between the human joints represented in the SMPL format and the joints of the target robot. For instance, both the AIST++ and PhantomDance data sets provide poses in the SMPL format. These poses can be converted into 24 joints, and our method can then be applied to a full humanoid robot.

The UR3 is a 6-DOF robotic arm, hence we use the right arm poses within each full human pose. To convert a human arm pose to the UR3 joints, we simplify the problem and reduce the 6 joints to 4. We consider two joints connecting the shoulder to the trunk, one at the elbow connecting the arm to the forearm, and one at the wrist connecting the hand to the forearm. This choice aligns with the first four joints of the UR3. Although we could explore incorporating hand rotation around the wrist, the inherent challenge in accurately detecting this in the data set led us to prioritise avoiding potential errors that could affect convergence.

To derive joint angles, we rely on the 3D keypoint annotations. Acknowledging that the arm's position is influenced by the movement of the rest of the body, we do transformations to maintain a fixed position of the base of the arm across different poses. We compute the unit vector between the two shoulder keypoints. Then, we determine the rotation required to align this vector parallel to the floor, and apply it to all keypoints. Subsequently, we find the midpoint between the shoulders, calculate the unit vector to the pelvis point, and apply a rotation making this vector perpendicular to the floor, effectively straightening the spine and maintaining consistent arm positioning relative to the trunk.

To obtain the joint angles, we calculate unit vectors between the shoulder and elbow, elbow and wrist, and wrist and the last point of the index finger. The first two joint angles correspond to rotations between the forearm unit vector and the vertical and outward-pointing axes from the human, respectively. The third joint angle accounts for the rotation between the forearm and arm unit vectors, while the last joint angle reflects the rotation between the arm and hand unit vectors. This process is repeated for every pose in the data set. The values of the joint angles range from $-\pi$ to $\pi$. This process is represented in Fig. \ref{fig:keypoints}.

\subsection{Synchronization}

For each music-dance pair, we acquire a sequence of audio feature vectors $\{f_{t_f}\}_{t_f=1}^{T_f}$  by sampling at a rate of 60 times per second and a sequence of poses $\{p_{t_p}\}_{t_p=1}^{T_p}$ obtained from the data set annotations. Because the annotations were not generated with the exact frequency of 60 times per second, it is likely that $T_f \neq T_p$. To pair the elements from the two sequences, we will use the timestamps provided alongside the annotations $\{t_{t_p}\}_{t_p=1}^{T_p}$. We compute the timestamp of each feature vector as $t_{t_f} = t_f * 1/60$ and then associate it with the pose with the closest timestamp.


\section{Music to Dance Translation}

Automatic music-to-dance generation is often framed as a sequence modelling challenge.
We propose to expand from this sequence modelling view a propose to model the music to dance generation problem as a direct translation task. Where the model translates sequences from the source language of audio to the sequences of the target language of dance. 
We propose two variants to perform music-to-dance translation, the first uses the Transformer architecture, while the second uses the Mamba architecture.

\subsection{Transformer}

While other methods have employed Transformers for music-to-dance generation \cite{aichoreographer}, they often under-utilise the Transformer's full capabilities. They frequently overlook crucial components such as the decoder, masked attention, or positional encoding. Moreover, many existing approaches rely on multiple networks trained with adversarial learning, which can be highly data-inefficient \cite{danceformer}. 
Additionally, some methods \cite{bailando} require the discretisation of the dance space, which may introduce limitations in expressiveness and precision.
In contrast, our work applies the translation framework to music-to-dance generation, harnessing the full capabilities of a single Transformer model.

Unlike typical natural language translation tasks where the entire source language sentence conditions the encoder, we face a constraint. Due to the extensive length of music pieces, ranging up to 8000 audio features, conditioning the encoder on the entire music piece is impractical. Consequently, we condition the encoder on a sequence of size $K$ of preceding audio feature vectors $\textbf{s}$. During batch data sampling, a music piece is randomly selected, followed by the choice of a trajectory index $t \in [0, T - 1]$, where $T$ is the length of the music piece.
The $K-1$ audio features preceding $t$ compose the sequence. The sequence is padded with zero-filled vectors to the right, if its length $L_s < K$. This sequence is paired with a padding mask, creating the input of the encoder. This mask is a binary sequence of length $K$ with zeros in indexes where the audio feature vector was padded, and prevents the encoder from considering the padding elements.

On the decoder's side, after selecting index $t$, we sample the corresponding poses to the sampled audio features, but right-shifted by one. 
Akin to the start token in natural language processing, we append a padding pose of zeros to the start of the pose sequence. 
This shift conditions the encoder on past poses and prevents it from seeing the ground-truth. Similar to the audio features, this pose sequence is right-padded to reach a length of $K$ elements. The decoder also receives a causal binary masking matrix where the values under the diagonal are ones and the remaining are zero. This ensures that attention scores consider only past tokens, preventing access to future information.

Both audio feature vectors and pose vectors pass through individual linear layers, converting them to 128-dimensional vectors. To convey positional information, an embedding layer processes the index $t$ of each element in the $K$ sequence. The vocabulary's size is set to the length of the largest music piece in the data set plus one for the start token. The resulting embeddings are summed to both the audio and pose embeddings before entering the Transformer.

We employ a Transformer architecture where the encoder and the decoder are composed of 6 blocks, 8 attention heads, a dropout rate of 0.1, and feed-forward layers with 2048 neurons, and output 128-dimensional embeddings. The embeddings returned by the decoder pass through a final feed-forward layer to project them to four joint angles. Subsequently, a hyperbolic tangent activation function normalises the values between -1 and 1, which are then multiplied by $\pi$ to scale them to the range of the joint angles.

The Transformer's objective is to translate audio sequences into corresponding poses after learning the mapping from music-dance pairs in the data set. Given input sequences of audio features $\textbf{s}$ and right-shifted poses $\textbf{a'}$, the Transformer aims to output the non-shifted pose sequence $\textbf{a}$. The loss function is the L2 loss between the predicted joint angles and the ground-truth.



\subsection{Mamba}

The Mamba variant is much simpler than the Transformer. Mainly because it is composed of the Mamba backbone instead of an encoder or decoder which streamlines the processing. This means that the causal binary mask of the Transformer is not needed. Additionally, because Mamba keeps track of the previous state, it also does not require positional embeddings. Hence, the audio feature vectors and pose feature vectors pass through the respective embedding layers and the outputs are fed to the Mamba blocks directly. 
We employ a Mamba architecture composed 6 Mamba blocks, a dropout rate of 0.1, and feed-forward layers with 2048 neurons, and output 128-dimensional embeddings.
The embeddings returned by the Mamba blocks are fed through the same pipeline as previously described to obtain the joint angles. The loss function is also the L2 loss between predicted joint angles and ground-truth.

\section{Experiments}

\subsection{Experimental Setup}

We apply the \methodname{} on the Universal Robots 3 (UR3), a 6-DOF robotic arm.
Our experimental process begins with the application of the data processing pipeline (see Section \ref{sec:pipeline}).
This process yields a refined data set $D'$ of pairs of sequences of 438-dimensional audio features and 4-dimensional poses. The processed AIST++ data set comprises 60 pairs, organised into 10 music genres, each containing 6 pairs. The PhantomDance data set is comprised of 260 pairs. In all the experiments we compare the performance of our method using the Transformer or the Mamba architecture.

We made an ablation study to determine an adequate hyper-parameter configuration for the model with either the Transformer or Mamba variant. We varied the number of Transformer or Mamba layers, the size of the embeddings within the model, and the length of the token sequence upon which the model is conditioned. Subsequently, we evaluate the models based on the Average Joint Error (AJE) observed during training, and the model's size. All models are trained using the complete AIST++ data set.
The configuration that yields the lowest pair of AJE and model size for the \methodname{}-T variant entails 6 Transformer layers, an embedding size of 128, and a sequence length of 20. In contrast, for the \methodname{}-M variant, the sequence length is extended to 120. We will use these two architectures throughout the remaining experiments.

We evaluate \methodname{} on both the entire data set and individual music genres. In the case of single-genre experiments, our approach involves training \methodname{} on 5 data pairs within a specific genre and subsequently evaluating its performance on the remaining pair. To ensure robustness and generalisation, we perform cross-validation across the 6 possible combinations within each genre, calculating both mean and standard deviation metrics across these variations.

For experiments encompassing the complete data set, we adopt a validation strategy by assigning one pair per genre for validation purposes, leaving the remaining 5 pairs for training. Employing a similar cross-validation methodology, we iterate this process 6 times, presenting the aggregated mean and standard deviations over these runs.
This evaluation strategy enables us to gauge \methodname{}'s performance not only across diverse genres but also its ability to generate genre-coherent music, providing valuable insights into its robustness and generalisation capabilities.
Lastly, we evaluate the performance of our method on the PhantomDance data set by performing cross-validation across 10 different combinations of 26 pairs assigned to the validation set. 

We compare the performance of \methodname{}-T with \methodname{}-M. Unfortunately, direct comparison with previous music-to-dance approaches was not possible for two main reasons.
Firstly, many of the other methods are not readily available or accessible. Secondly, other methods produce different outputs such as velocities, key-poses, and Bezier curves. To compare with these methods, we would need to execute the velocities or curves on the robot and then measure the resulting joint positions. Additionally, this process requires converting our data set to match the velocities or curves format. This conversion poses challenges, including determining the appropriate velocities/curves that accurately represent the original human movements and also correspond to the robot's execution, adding further complexity to the comparison process.

Each experiment comprises a 250k update steps, utilising a batch size of 16 and a learning rate of $10^{-4}$. Model validation is conducted at regular intervals of 1k updates. During validation we iterate over each music in the music-dance pairs and use the model to generate a sequence of poses. This sequence of poses is then compared with the ground-truth poses using the following two metrics:
\begin{itemize}
    \item \textbf{Average Joint Error (AJE)} : Average joint error in radians between translated poses and the ground-truth for a given music sequence $AJE(p_i^G, p_i^T) \in [0, \infty[$. 
    \item  \textbf{Fréchet Inception Distance (FID) \cite{fid}} : Evaluates how close the distribution of generated dances is to that of the ground-truth dances.
\end{itemize}

\subsection{AIST++ Music Genre Generalisation}

\begin{table}[t]
\centering
\caption{FID and AJE metrics on AIST++ data set.}
\label{tab:aist}
\begin{tabular}{c|cc|cc}
\multirow{2}{*}{\textbf{\begin{tabular}[c]{@{}c@{}}AIST++\\ Genre\end{tabular}}} & \multicolumn{2}{c|}{\textbf{\methodname{}-T}}       & \multicolumn{2}{c}{\textbf{\methodname{}-M}}          \\ \cline{2-5} 
                                                                                 & \textbf{AJE (rad)}     & \textbf{FID (\%)}      & \textbf{AJE (rad)}   & \textbf{FID (\%)}    \\ \hline
All Genres                                                                       & 0.58 +/- 0.10          & 1.32 +/- 0.64          & \textbf{0.57 + 0.06} & \textbf{0.96 + 0.31} \\
mBR                                                                              & 0.57 +/- 0.13          & 0.46 +/- 0.25          & \textbf{0.53 + 0.04} & \textbf{0.40 + 0.18} \\
mHO                                                                              & \textbf{0.25 +/- 0.03} & \textbf{0.06 +/- 0.04} & \textbf{0.25 + 0.04} & 0.13 + 0.08          \\
mJB                                                                              & \textbf{0.57 +/- 0.16} & 0.79 +/- 0.75          & \textbf{0.57 + 0.15} & \textbf{0.61 + 0.50} \\
mJS                                                                              & 0.35 +/- 0.13          & 0.38 +/- 0.36          & \textbf{0.34 + 0.08} & \textbf{0.22 + 0.19} \\
mKR                                                                              & \textbf{0.58 +/- 0.27} & \textbf{0.76 +/- 0.59} & 0.61 + 0.19          & 0.91 + 1.05          \\
mLH                                                                              & 0.30 +/- 0.04          & 0.10 +/- 0.05          & \textbf{0.29 + 0.05} & \textbf{0.06 + 0.01} \\
mLO                                                                              & \textbf{0.45 +/- 0.18} & 0.74 +/- 0.92          & 0.47 + 0.14          & \textbf{0.51 + 0.65} \\
mMH                                                                              & \textbf{0.33 +/- 0.10} & \textbf{0.11 +/- 0.06} & \textbf{0.33 + 0.05} & 0.12 + 0.05          \\
mPO                                                                              & \textbf{0.23 +/- 0.06} & \textbf{0.16 +/- 0.11} & 0.29 + 0.10          & 0.24 + 0.25          \\
mWA                                                                              & \textbf{0.62 +/- 0.28} & \textbf{0.75 +/- 0.96} & 0.63 + 0.25          & 1.03 + 1.68         
\end{tabular}
\end{table}

We investigate the performance of the two variants using the processed AIST++ data set. 
We evaluated on individual music genres and also using all music genres.
The results, showcasing the mean and standard deviations after cross-validation of AJE and FID on the each genre's validation set, are presented in Table \ref{tab:aist}.
The results show small differences between the two variants with the Transformer variant offering a slightly lower average AJE and FID across the experiments. The model seems to perform the best on the 'mPO' music genre and the worst on the 'mWA'. Overall \methodname{}-T offers an average AJE of 0.44 radians and an average FID of 0.51\% across the 11 experiments. These low values validate the model's ability to successfully learn the mapping between the audio and dance languages and consequently generalise to unseen audio pieces. 

\subsection{PhantomDance Evaluation}

\begin{table}[t]
\centering
\caption{FID and AJE metrics on PhantomDance data set.}
\label{tab:phantom}
\begin{tabular}{cc|cc}
\multicolumn{2}{c|}{\textbf{\methodname{}-T}} & \multicolumn{2}{c}{\textbf{\methodname{}-M}} \\ \hline
\textbf{AJE}         & \textbf{FID}       & \textbf{AJE}     & \textbf{FID}    \\ \hline
0.87 $\pm$ 0.02        & 0.39 $\pm$ 0.02        & 0.73 $\pm$ 0.01                & 0.82 $\pm$ 0.01             
\end{tabular}
\end{table}

Finally, we assess performance of the two variants using the processed PhantomDance data set. The summarised results are presented in Table \ref{tab:phantom}. Each value is the mean and standard deviation of cross-validation across 10 combinations of 90/10 train/validation splits. Notably, both AJE and FID metrics exhibit an increase compared to the previous results obtained from the AIST++ data set. This is attributed to the augmented complexity of the data set. Specifically, the choreographies within PhantomDance are lengthier and exhibit greater diversity as opposed to AIST++ where the pairs can be organised by music genre. Nevertheless, the model consistently achieves significantly lower AJE compared to a random model, along with a low FID score. This demonstrates the model's capability to learn the mapping from music to dance, thus validating our approach of framing music-to-dance generation as a language translation task.

\section{Conclusions}

We propose to model the music-to-dance generation task as a language translation problem. We describe two variants: \methodname{}-T and \methodname{}-M using the Transformer and the Mamba architecture, respectively.
We researched previous music-to-dance methods which validated the selection of audio features, and presented an approach for mapping human poses from keypoints to joint angles of a robotic arm. \methodname{} can also be applied to humanoid agents.

Evaluation on music-dance pairs from the AIST++ and PhantomDance data sets, through AJE and FID metrics, demonstrates that \methodname{} can robustly and efficiently translate diverse and unseen music to high-quality dance motions coherent within the genre. 

\section*{Acknowledgments}

This work is supported by NOVA LINCS ref. UIDB/04516/2020 (https://doi.org/10.54499/UIDB/04516/2020) and ref. UIDP/04516/2020 (https://doi.org/10.54499/UIDP/04516/2020) with the financial support of FCT.IP, and also through the research grant 2022.14197.BD.

\end{document}